# Modelling of pedestrian movement around 90° and 180° bends

*Bernhard Steffen, Armin Seyfried*
*Jülich Supercomputing Centre, Forschungszentrum Jülich GmbH,*
*Wilhelm-Johnen Strasse, 52425 Jülich, Germany*

## ABSTRACT

For the planning of large pedestrian facilities, the movement of pedestrians in various situations has to be modelled. These include different densities and different usages, at least standard operation mode and evacuation. Many tools for pedestrian planning are based on cellular automata (CA), discrete in space and time.

It is common experience that CA have problems with modelling sharp bends in wide corridors and especially with 180° turns as they are frequent in staircases. They tend to move the pedestrians to the innermost lanes far too strongly, thereby reducing the capacity. In the extreme, only a small fraction of the width will be used after the bend. There have been some remedies proposed, but a systematic investigation is lacking, and there is no accord on the causes of the problem.

In the paper, we present a comparison of various implementations of CA - for treating 90° and 180° bends with different width. We compare different models for the static floor field which provides the persons orientation. We also indicate how a density dependant static floor field can give a much increased capacity while still allowing single walkers to cut corners. We test at which position the walkers are most likely to get stuck and relate this to differences between CA models and real-world behaviour.

The behaviour in the model is compared with observation of people walking through a 90° bend in a corridor and walking down a wide staircase. It is shown that with a proper floor field, at least the average behaviour of people can be simulated. Capturing the fluctuations present in reality is beyond the capacity of the simple models investigated.

## INTRODUCTION

The last century has seen a tremendous growth of the world's population, together with an increasing concentration of the population in big cities. Therefore, the safe and effective organisation of pedestrian facilities has gained in importance, as the crowds gathering in one place (for ceremonies, entertainment, work or education) get more frequent and larger.

Basic rules for planning of pedestrian facilities were obviously known to Greeks and Romans, as the quality of the egress routes e.g. at the theatre in Ephesus or the Coliseum in Rome witnesses. Modern systematic planning began in the late 19[th] century with building regulations requiring a certain door width per number of people. These were motivated by some catastrophic fires in theatres and ballrooms and aimed exclusively at safe emergency egress [1]. As buildings - notably office buildings, which have much more complicated egress routes than facilities for entertainment - became bigger, detailed planning became necessary, where the building egress was modelled as a tree or network of pathways that take time, but have no active capacity restriction, and nodes (doors, junction of floors, etc), that do have active capacity restrictions [2, 3, 4]. These models may be tedious to set up and use, but they allow reliable hand calculations of the time needed to clear a building and give predictions on critical situations during evacuation. From building regulations and the performance based egress analysis it follows that turns are not considered a problem there. This may be a reason that no empirical data is available.

Starting at ~1990, a modelling of the individual movement of all persons in a building was attempted, and the macroscopic picture became the sum of many local movements obeying simple rules. There are two basic approaches: the cellular automata (CA) model where space and time are discrete and people are moving from one space element to another one according to some transition rules, and models in continuous space where persons are moved according to Newton's laws by forces generated mostly internally as a reaction to the desired momentary destination and the local environment. While these are easier to use than the earlier models - the time spend in pathways and the capacity of bottlenecks needs not be specified, it comes as a result - the definition of the rules or forces is much more demanding, because small errors there may build up to large errors in output quantities. There is up to now no full understanding how subtle differences in the modelling change the outcome, so these models have to be tested further against each other and against reality in various situations. Part of this has been done in [5]. We will test two further simple situations here - the walking around a bend in a wide (>2m) corridor and the turning at the landing of a ramp or staircase – because some simulations presented, unfortunately all unpublished - did not look very realistic. Our aim is not to give the final answer to this modelling, but to investigate how changes in guiding the persons in the simulation influence the outcome.

## 1 CELLULAR AUTOMATA

Cellular automata are the most widely used approach. The commercial code PedGo [6] is a CA code; others (e.g. buildingEXODUS [7]) are based on rules for discrete hopping related to CA. They have demonstrated their ability to give good estimates of evacuation times of high rise buildings [5], while details may still need improvement. While they differ in many aspects, the basics are identical.

## 1.1 Introduction to the general theory

The principles of cellular automata for simulating pedestrians are explained in many places, e.g. [8,9]. For readability, we repeat here the description from [9]. The floor geometry is discretised into tiles, usually of 40cm by 40cm size. An initial distribution of persons on the tiles is defined. In every time step each person can move to another tile (or stay were he is) according to a probability depending on

- The availability of free space (only one person per tile at any time).
- A (static) floor field describing the intended direction.
- Possibly a (dynamic) floor field summarising the movement on the tile during the last few time steps.
- Personal data, describing e.g. handicaps.

With a time step of $\approx$ 0.3s this gives a reasonable speed of free movement of about 1.3m/s. The static floor field S is usually (we will investigate a different construction, too) derived from the gradient of the distance to the destination (exit) in some metric, the Manhattan metric being the most common one because it is extremely simple to implement even in complicated geometries. With this, the distance is simply the minimal number of tiles the person has to pass before reaching the destination, and $S_{ij}$ for neighbouring cells can only take values -1, 0 and 1. For a person on tile i, the probability of movement to a neighbouring cell j is given by

$$p_{ij} = N_i \exp(k_S S_{ij} + k_D D_{ij})(1-n_j)\chi_{ij} \qquad (1)$$

- $n_j$ is the occupation number $n_j = 0|1$ ,
- $\chi_{ij}$ gives reachability of tile j from tile i, $\chi_{ij} = 0|1$ ,
- $S_{ij}$ is the static floor field weighted with a sensitivity parameter $k_S$ that may depend on the person on tile i.
- $D_{ij}$ is the dynamic floor field weighted with a sensitivity parameter $k_D$, $k_D = 0$ here.
- $N_i$ is the normalisation factor such that the sum of all probabilities for one person is 1:
  $N_i = 1/\sum_j (\exp(k_D D_{ij} + k_S S_{ij})(1-n_j)\chi_{ij})$ .

The time step update may either be done in parallel with some conflict resolution, or sequential. To avoid directional bias and other artefacts in the sequential update, a random order changing from step to step is strongly recommended. The reachability factor $\chi_{ij}$ serves two purposes. It describes obstacles and walls as well as the choice of neighbourhood (Fig1). The most common choice is the Neumann neighbourhood with 4 direct next neighbours, the Moore neighbourhood with 8 neighbours (diagonals included) is less common, others - stepping over more than on cell in one time step – have also been investigated [10], but are not used in production.

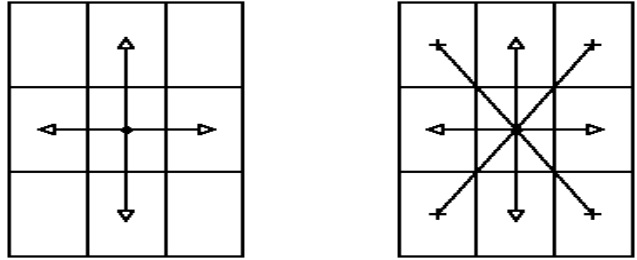

**Fig 1**: Places reachable with 4 and 8 direct neighbours

## 1.2  The influence of the static floor field

As we only want to investigate the influence of the static floor field on the dynamics of walking around a 90° or 180° turn, we keep the method as simple as possible otherwise, that is: No dynamic floor field, only the four direct neighbour cells are reachable, $k_s = 4$ for all persons identical, random sequential update. Four different static floor fields $S_{ij}$ are compared:
- F1: derived from the Manhattan distance,
- F2: derived from the Euclidian distance,
- F3: a field keeping the distance from the inner wall constant,
- F4:  a* F1 + (1-a) F3 with a=1/4.

For a simple straight corridor all fields are identical, and the capacity of such a corridor was taken as reference. From the results of our studies a width of 5 tiles and a length of the inner walls of twice the width were determined as critical. This shows all the effect of the floor fields, while with a smaller width these much are less pronounced.  Actually, for width one all floor fields are identical, and for width two the difference is not more than the normal scatter of the results. The measurements were arranged such that the space was filled quite dense, then the walking was started. Every person reaching the end of the corridor was entered at the beginning again in a random position if there was space available, if not he was assigned a waiting position. After a short initial phase, long enough to reach a stationary state, the measurement was started. Tests showed that the number of persons inside the corridor stayed fairly constant, and that the flow obtained thus increased monotonously with the number of persons initialised up to a maximum depending on the floor field chosen.

|  |  | Width of 5 tiles | | Width of 10 tiles | |
|---|---|---|---|---|---|
| Corridor type | Static field type | Flow per time step | Pers. Waiting outside | Flow per time step | Pers. waiting outside |
| straight | F1 | 2.14 | 2.6 | 3.99 | 11.5 |
|  | F1 | 0.98 | 2.2 | 1.25 | 16.5 |
| one left turn | F2 | 0.81 | 1.9 | 1.20 | 5.1 |
|  | F3 | 1.98 | 7.2 | 3.98 | 28.2 |
|  | F4 | 1.97 | 4.7 | 3.65 | 22.3 |
| two left turn | F1 | 0.75 | 13.7 | .78 | 48.7 |
|  | F2 | 0.84 | 12.0 | 0.92 | 56.8 |
|  | F3 | 1.91 | 15.4 | 3.80 | 26.1 |
|  | F4 | 1.81 | 17.3 | 3.17 | 57.3 |

**Table 1**  Simulation Results for different setups of CA models

Table 1 gives the results for maximum flow for all fields. This shows that with floor fields F1 and F2, the turns are strong bottlenecks. Even worse, a wider floor has hardly more capacity. This is definitely not in agreement with everyday observations, even if they cannot be quantified. With floor field F3, the capacity of the turn is more than 90% of the capacity of a straight corridor, which may be true, but seems optimistic. Floor field F4 shows that a linear combination of field F3 and field F1 can be tuned to give the proper result.

Beyond the capacity figures, the simulation can analyse the movement patterns. For a 90° turn and floor fields F1 and F3 we show a typical snapshot and the distribution of occupation and movement over time.  From the snapshot (Fig. 2) it can be seen that  for field F1 the inner lanes before the first curve are blocked by persons turning in from outer lanes (the result from field F2 looks almost the same), while with field F3 the outer lanes take a wider curve and the inner lanes move fairly un-obstructed.

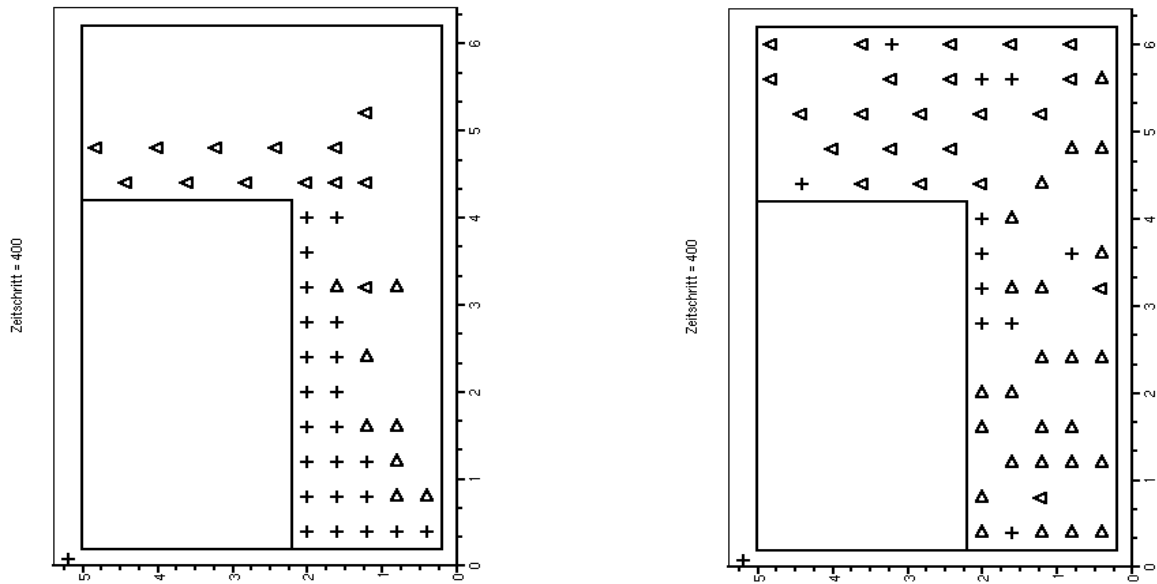

**Fig. 2** Shnapshots of movement with field F1 (left) and F3 (right).
+ indicates standing, an arrowhead indicates motion

There is always a small percentage of people not moving, this is typical for CA modelling as the probability for not moving is $1/N_i$, which depends on $K_S$. It is also clear from the procedure of refeeding people into the system that the cells at the entrance are always occupied and therefore there is not much free movement near the entrance.

# 1  CONTINUOUS SPACE MODELS

The description of pedestrians as (mostly) self-driven particles walking in continuous space and time due to forces they generate themselves from the desire to reach some destination and to avoid collisions (plus forces from pushing in very dense crowds) can give more realistic individual movements than CA methods, but are much more demanding and (probably as a consequence) not as widely used. There are completely different proposals for the specification of these forces [11, 12], and the models are mostly in an experimental state.

## 1.1 The social force model

The social force model (SFM,[11,13,14)is the oldest and most matured of the SDP models, and it will be used here. In its original version it is intended for qualitative rather than quantitative phenomena and thus gave some inside into self-organisation of pedestrian flows rather than numbers on evacuation times. This is sufficient for the present purpose. It has been developed into a quantitative method [15], but this is a commercial product and details are not available.

The social force model assumes a Force F as sum of an accelerating force $F_a$ proportional to the difference between the desired and the present velocity, an influence of other people $F_r$ given by the sum of binary person-person interactions $F_{i,j}$ which are derived from an exponential potential, plus wall forces $F_W$ similar to $F_r$. The simplest (original) form depends on distance only.

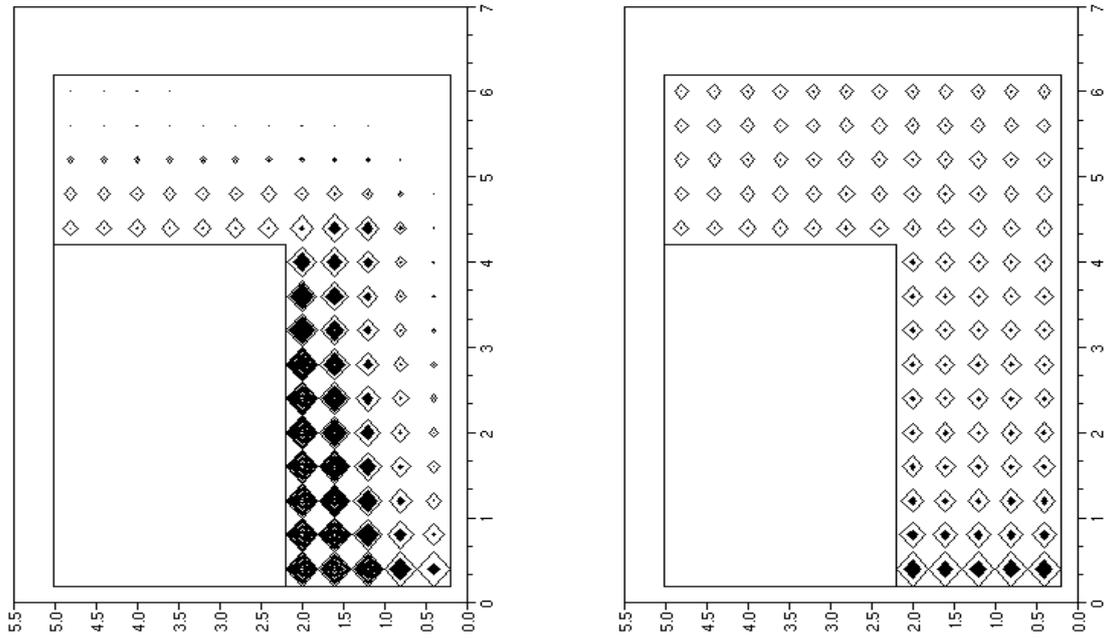

**Fig. 3** Summary of movement with field F1 (left) and F3 (right).
Open diamonds give occupation, closed diamonds inside occupation by people standing still

$$F = F_a + F_r + F_w; \tag{2}$$

$$F_a = (v_0-v)/\alpha, \quad v_0 \text{ the desired velocity,} \quad \alpha \approx 1.6 \tag{3}$$

$$F_r = \sum_j F_{i,j}; \quad F_{i,j} = C \exp(-\lambda \| x_i - x_j \|), \text{ here } \lambda = 0.3, C = 7.3 \tag{4}$$

It gives a decent agreement for the trajectories, but lane formation is much too strong.

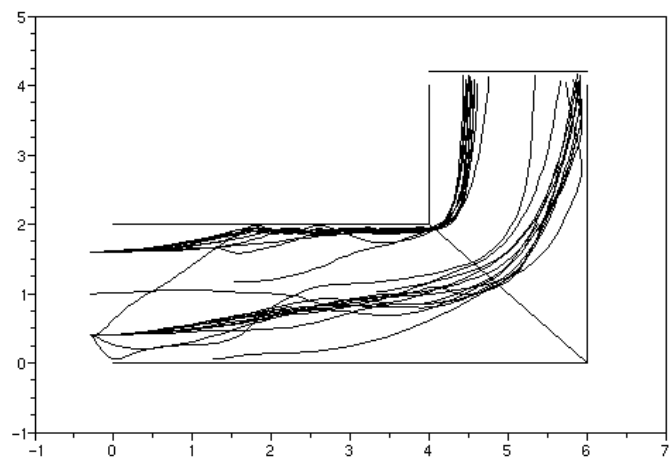

**Fig. 4** Tracks from Social force model with one intermediate and one final destination line

## 2     EXPERIMENTAL FINDINGS

Experiments have been performed for various situations within the context of the Hermes project [16] just recently and will be analysed in detail, but at the moment only summery data is available. The experiments have been recorded by stereo wide angle overhead cameras, and the transformation of the raw camera data into trajectories needs some effort. There has been some recording with standard cameras, too. From these, the software PetTrack [17] can produce trajectories of moderate accuracy using the (invalid) assumption that all persons are of the same size. However, these cameras cover only a small part of the floor, so they do not give all the data needed for a good comparison.

### 3.1 Walking round a corner

The experiments have been recorded on video using two pairs of wide angle stereo cameras which together cover the total region and one standard camera covering the central region only. Stereo cameras are used such that the height of people's heads is known and thus accurate trajectories are obtained. Otherwise the perspective distortion cannot be handled correctly. Unfortunately, the stereo camera pictures come in a non-standard data format and have not been transformed yet, so only the standard camera data is available, and only for one (the most interesting) width (2.4m). All data will be available by the end of 2009. For calculation of positions we assumed everybody to be 180 cm tall, which introduces positional errors of about 10 cm. The trajectories show that the people cut the corner a bit, but behind the corner they basically regain their former distance from the inner wall. Grouping these distances by cells of 40 cm width, the numbers for the cells from inner to outer were 0, 14, 24, 18, 7, 1 going in and 0, 24, 21, 14, 5, 0 going out. This is a weak tendency for moving to the inside, similar to the CA model with floor field F4. On the other hand, the fact that the innermost lane is not used at all, which may have to do with the fact that in case of bidirectional traffic this would be dangerous at the corner, is not captured by any floor field.

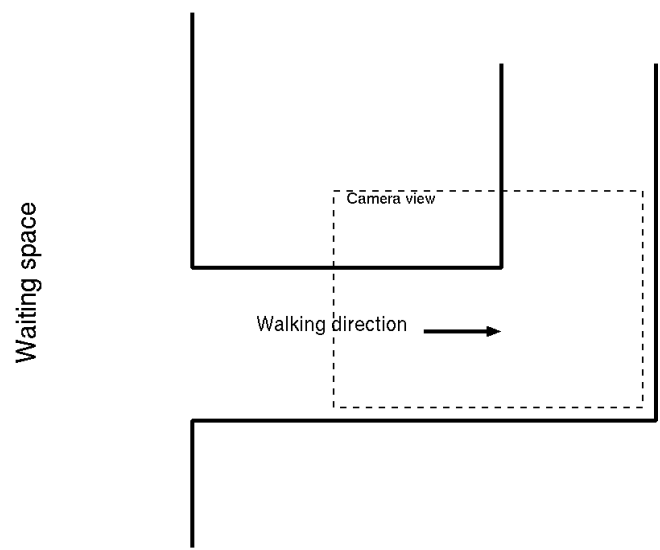

**Fig. 4** Experimental setup for walking round a corner

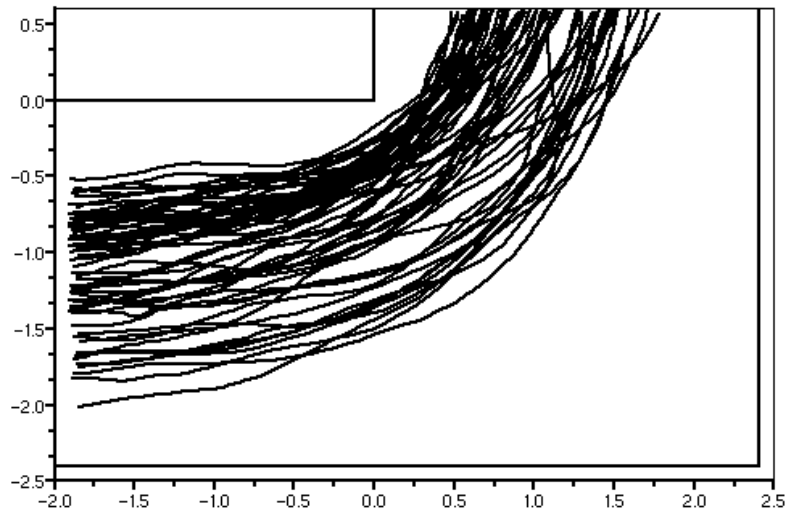

**Fig 5.** Trajectories of people walking round a corner (the walls are longer than shown)

Obviously, the corridor is not used to capacity, and everybody keeps away from the walls. The average flux is 1.25 persons per second. From other experiments [18] we know that the entrance to a corridor is the actual bottleneck, while over the length of the corridor speed increases and density drops. The effect of the 90° turn is obviously too small to fill the corridor again.

## 3.2 Walking down a staircase

The experiments on a staircase of 2.4 m width were documented on video, but for this the calculation of positions from the video is not available yet. Therefore, only the most obvious observations are reported. The experiments differ from the CA calculations in that the landings are not rectangular but semi-circular. Again, the entering of the staircase was the problem, afterwards the density decreased and the descend was almost unobstructed. Even three stories down, after 5 turns of 180°, people would occasionally walk four abreast, even though on average the inner part of the stairs was preferred. There was no indication of any obstruction caused by the turning. This may, however, be due to the fact that the stairs have lower specific capacity than the level ground, and additionally the entering of the stairs is even more restricted, so that a moderate obstruction would pass unnoticed. Walking upstairs shows the same picture, the turn at the landings being no bottleneck. Here the capacity of the staircase is even more restricted.

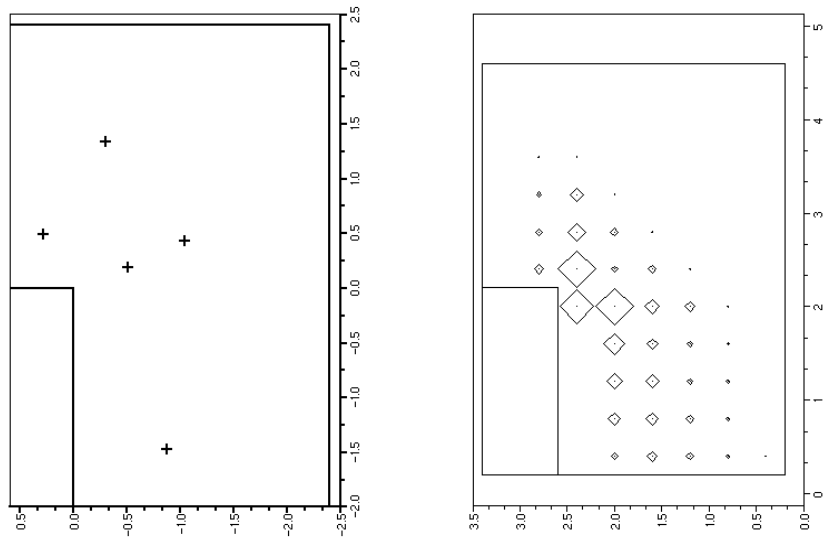

**Fig. 6** Snapshot and summary of cell occupation for experiment

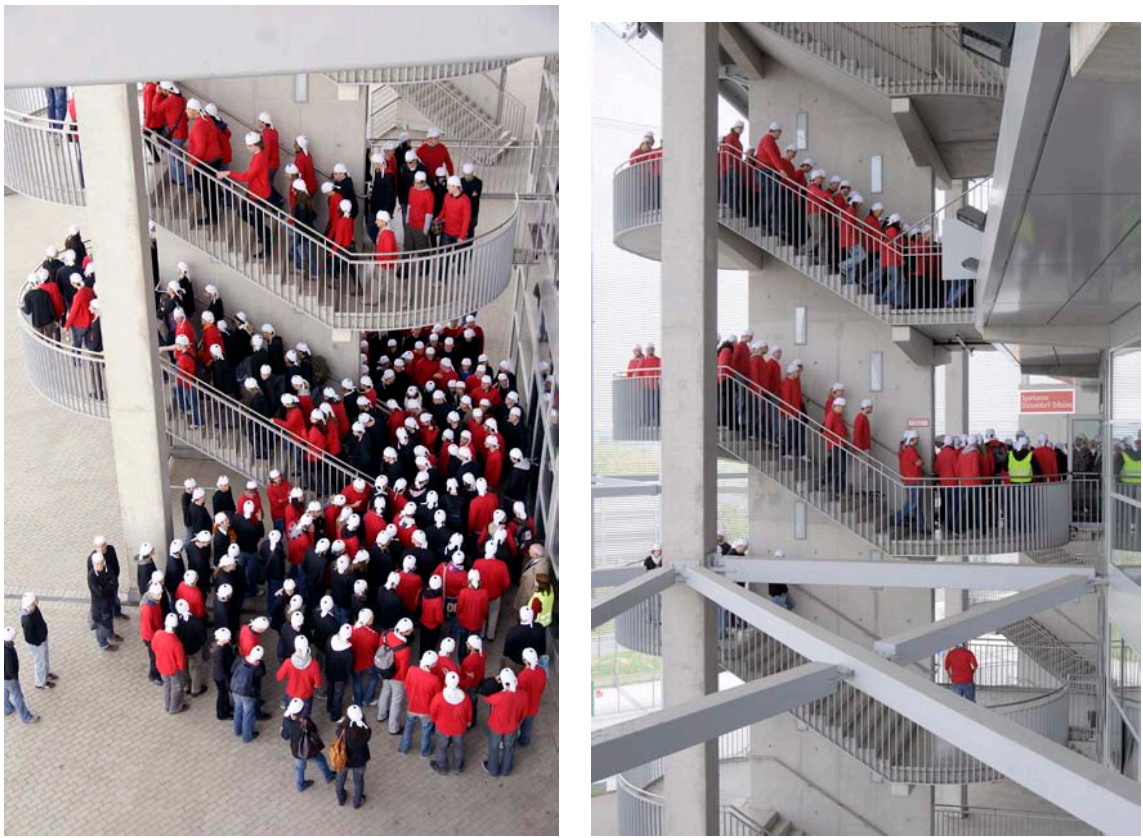

**Fig. 4** Experiments walking up and down the staircase

## 4 CONCLUSIONS

For the situation inspected here, the conventional floor field for a cellular automata approach makes 90° corners in wide floors appear as bottlenecks for the egress. This feature is just visible for floors 3 lanes (1.2m) wide and becomes more pronounced and quite unrealistic for floors wider than 2m. A small modification of the floor field, the addition of a wall parallel field, changes this behaviour such that bends have an egress capacity not much lower than straight corridors of the same width. This change does not increase the computing time or complexity. There are other attempts to solve the same problem, like introducing dynamic floor fields [9] or some local optimization [17] to people, but these add considerably to the computational complexity.

## ACKNOWLEDGMENTS

We thank M. Boltes for the pictures and the trajectories from the experiment.